\documentclass[12pt,preprint]{emulateapj}

\usepackage{graphicx}
\usepackage{txfonts}
\usepackage{threeparttable}

\shorttitle{Revisiting the Gamma-Ray Burst Classification}
\shortauthors{Zhang et al.}

\begin{document}

\title{Revisiting the Long/Soft $-$ Short/Hard Classification of Gamma-Ray Bursts in the Fermi Era}

\author{\sc Fu-Wen Zhang\altaffilmark{1,2,3}, Lang Shao\altaffilmark{4}, Jing-Zhi Yan\altaffilmark{1,2}, and Da-Ming Wei\altaffilmark{1,2}}
\altaffiltext{1}{Purple Mountain Observatory, Chinese Academy of Sciences, Nanjing 210008, China;}
\altaffiltext{2}{Key Laboratory of Dark Matter and Space Astronomy, Chinese Academy of Sciences, Nanjing 210008, China;}
\altaffiltext{3}{College of Science, Guilin University of Technology, Guilin 541004, China;}
\altaffiltext{4}{Department of Physics, Hebei Normal University, Shijiazhuang 050016, China.}
\email{fwzhang@pmo.ac.cn(F.-W.Z.)}

\begin{abstract}

We perform a statistical analysis of the temporal and spectral
properties of the latest Fermi gamma-ray bursts (GRBs) to revisit
the classification of GRBs. We find that the bimodalities of
duration and the energy ratio ($E_{\mathrm{peak}}$/Fluence) and the
anti-correlation between spectral hardness (hardness ratio ($HR$),
peak energy and spectral index) and duration ($T_{90}$) support the
long/soft $-$ short/hard classification scheme for Fermi GRBs. The
$HR - T_{90}$ anti-correlation strongly depends upon the spectral
shape of GRBs and energy bands, and the bursts with the curved
spectra in the typical BATSE energy bands show a tighter
anti-correlation than those with the power-law spectra in the
typical BAT energy bands. This might explain why the $HR - T_{90}$
correlation is not evident for those GRB samples detected by
instruments like {\it Swift } with a narrower/softer energy bandpass. We also analyze
the intrinsic energy correlation for the GRBs with measured redshifts
and well defined peak energies. The current sample suggests
$E_{\mathrm{p,rest}}=2455\times (E_{\mathrm{iso}}/10^{52})^{0.59}$
for short GRBs, significantly different from that for long GRBs.
However, both the long and short GRBs comply with the same
$E_{\mathrm{p,rest}}-L_{\mathrm{iso}}$ correlation.

\end{abstract}


\section{INTRODUCTION} \label{sec:intro}

The field of gamma-ray bursts (GRBs) has rapidly advanced in recent
years especially after the launch of NASA missions Swift (in 2004,
Gehrels et al. 2004) and Fermi (in 2008, Atwood et al. 2009). A
physical classification of GRBs is still a basic open question
(e.g., Zhang 2011). According to the traditional classification
schemes, GRBs can be divided into long and short ones, based on the
well-known bimodal distribution of their durations monitored by the
Burst And Transient Source Experiment (BATSE, Meegan et al. 1992),
which also show different spectral hardness ratios ($HR$;
Kouveliotou et al. 1993). The $HR$ in conjunction with the duration
provides a means for classification, e.g., the long/soft class
comprises roughly 3/4 of the population and the short/hard class
comprises the other 1/4 for BATSE GRBs (Kouveliotou et al. 1993; Qin
et al. 2000). The difference between long and short GRBs is further
established by the observations of afterglows and host galaxies.
The fact that several nearby long GRBs are associated with
Type Ic Supernovae (SNe) and most long GRB host galaxies are found
be dwarf star-forming galaxies favors the speculation that most long
GRBs are accompanied by massive stellar explosions (see Woosley \&
Bloom 2006, for a review). Some nearby short GRBs (or short GRBs
with a long-soft extended emission) have host galaxies that are
elliptical or of early type, with little star formation (Fox et al.
2005; Gehrels et al. 2005; Berger et al. 2005a; Barthelmy et al.
2005). This points towards a different type of progenitor, e.g.
compact object mergers (see Nakar 2007, for a review).

This dichotomous picture was soon challenged by some following
observations. GRB 060614 and GRB 060505 are two nearby long GRBs
that did not have bright SN associations, sharing similar properties
to short GRBs (Fynbo et al. 2006; Della Valle et al. 2006; Xu et al.
2009). Three high-$z$ GRBs 080913, 090423 and 090429B have
rest-frame durations shorter than 1 s, but are likely related to
massive stars\footnote{The other possibility that these high
redshift bursts may be from the super-conducting cosmic strings has
been tightly constrained (Wang et al. 2011).} (Greiner et al. 2009;
Salvaterra et al. 2009; Tanvir et al. 2009; Cucchiara et al. 2011).
An observed short GRB 090426 was found in many aspects similar to
long GRBs and it was also probably linked to the death of a massive
star (Antonelli et al. 2009; Levesque et al. 2010; Xin et al. 2011;
Th{\"o}ne et al. 2011; Nicuesa Guelbenzu et al. 2011). These results
suggest that certain observational properties (e.g., long vs. short
duration) do not always refer to certain types of progenitor (see
Zhang et al. 2009 and the references therein). Moreover, the bimodal
duration distribution was not presented in observation for some
detectors with a narrow/softer energy bandpass such as HETE-2 and
Swift, and the distinction of short/hard and long/soft in the
hardness-duration panel is not very clear (e.g., Sakamoto et al.
2011 and the references therein; Shao et al. 2011).
Recently, Guiriec et al. (2010) analyzed 3 very bright short
GRBs observed by the Fermi Gamma-Ray Burst Monitor (GBM, Meegan et
al. 2009), and found that short GRBs are very similar to long ones,
but with light curves contracted in time and with harder spectra
stretched toward higher energies. They also showed that the hardness
evolutions during the bursts follow their flux/intensity variations,
similar to long bursts. By studying the composite light curves
including both the prompt and afterglow emission of GRBs detected by
Swift, Shao et al. (2011) found the similarity of the radiative
features between the long and short bursts.  They also proposed that
the spectral evolution of the prompt emission might be an important
factor that determines the correlation between the hardness ratio
and duration.

Theoretically speaking, the duration may be not an unique indicator
for the physical category of a GRB. In the pre-Swift era, it was
already known that (1) the coalescence of two compact objects can
produce either long or short GRBs, depending on the remnant formed
in the merger (a short burst is likely if the remnant is a stellar
black hole while a long burst is expected if the remnant is a
super-massive neutron star; see Kluzniak \& Ruderman 1998 and the
references therein); (2) the collapse of a massive star usually
produces long GRBs but the possibility of generating short events
can not be ruled out (Zhang et al. 2003; Fan et al. 2005). One may
appeal to multiple observational criteria to judge the correct
physical category of the GRB progenitor model that is associated
with a certain GRB (e.g., Zhang et al. 2009; Fan et al. 2005; Kann
et al. 2011). Recently, Goldstein et al. (2010) found that
the distribution of the $E_{\mathrm{peak}}$/Fluence energy ratio has
a clear bimodality by analyzing the complete BATSE 5B Spectral
Catalog. An obvious distinction between long and short bursts
emerges from this bimodal distribution. This result was
further confirmed by an analysis of 382 GRBs from the GBM spectral
catalog (Goldstein et al. 2011). Another phenomenological
classification method for GRBs was proposed by L{\"u} et al. (2010).

It is urgent to discuss the classification issues further in the
Fermi era. The preliminary analysis results of GRBs observed
by Fermi-GBM presented by e.g., Bissaldi et al. (2011), Nava et al.
(2011, hereafter N11), and Shao et al. (2011) confirmed the duration
bimodality found in the BATSE data (Kouveliotou et al. 1993).
Whether the hardness-duration correlation is also consistent with
the BATSE result and the short/hard $-$ long/soft dichotomy
classification scheme is robust have not been explored yet.
In this paper, we present a systematic analysis of the
temporal and spectral properties of Fermi-GBM GRBs catalogued by N11
and analyze the intrinsic energy correlation of GRBs with measured
redshift and well defined peak energy. This paper is structured as
follows. The data and sample are presented in Section 2. In Section
3, we revisit the distribution of duration, the correlation between
hardness and duration, the distribution of energy ratio and the
intrinsic energy correlation. In Section 4, we give a summary.

\section{DATA AND SAMPLE} \label{sec:data}

The Fermi-GBM detected 438 GRBs by the end of March 2010.
The spectral properties of these GRBs have been analyzed and
published by N11. Out of the 432 GRBs for which it was possible to
perform the spectral analysis, 323 bursts have curved spectra which
are well fitted by the Band (Band et al. 1993) or by a
cut-off power law (CPL) model. The remaining 109 bursts are best
fitted with a simple power-law. In addition, the peak flux spectra
of 235 bursts could be extracted and fitted with a Band or CPL
model. The detailed data extraction and analysis can be
found in the catalog by N11, which we adopted in the following
analysis. Recently, the first two years Fermi-GBM GRB catalog was
released (see, Paciesas et al. 2012; Goldstein et al. 2012). The
duration ($T_{90}$) of 487 GRBs are reported in the catalog.
Combined with the duration data, we separate the 427 GRBs with an
analyzed spectrum into different samples (5 cases are not included
for lack of $T_{90}$ data): 322 GRBs with the curved spectra (Band
or CPL model) represent {\it sample 1}, while 103 GRBs with the
power-law spectra represent {\it sample 2}\footnote{The two GRBs
with unusually soft spectra ($\alpha=-3.52$ for GRB 100112 and
$\alpha=-3.98$ for GRB 100207) are excluded in the analysis.}.
Furthermore, the 234 GRBs with a fitted peak flux spectrum represent
{\it sample 3} (one burst without $T_{90}$ data is excluded).

To analyze the intrinsic energy correlation of the different GRB
classes, we collect the GRBs with known redshift and well defined
peak energy up to the end of May 2011. This sample includes 110 long
GRBs and 7 short GRBs detected by BeppoSAX, HETE-2, Swift, Suzaku
and Fermi. Most of data are taken from Amati et al. (2008, 2009),
Nava et al. (2008), Ghirlanda et al. (2009, 2010) and the references
therein, as well as the GRB Coordinates Network (GCN). The isotropic
gamma-ray energy ($E_{\mathrm{iso}}$) and luminosity
($L_{\mathrm{iso}}$) are calculated in the rest frame in the energy
range $1-10000$ keV.

\section{TEMPORAL AND SPECTRAL ANALYSIS} \label{sec:analysis}

\subsection{Duration and Hardness} \label{subsec:dur}

Distribution of duration is crucial for the traditional GRB
classification. We analyze the distribution of durations of Fermi
GRBs in N11 catalog in detail. As shown in Figure 1, the duration
distributions are bimodal either for all the bursts or for samples 1
and 2. The ``short" and ``long" GRBs are well separated in the
log-normal plot, which is consistent with the previous findings
(e.g., Bissaldi et al. 2011; N11; Shao et al. 2011). For all the
GRBs in N11, we find a cental value $\mu_1=-0.23$ (i.e.,
$T_{90}\sim0.59$~s; with a standard deviation $\sigma_1=0.47$) and
$\mu_2=1.42$ (i.e., $T_{90}\sim26.3$~s; with a standard deviation
$\sigma_2=0.47$) for short and long bursts, respectively. For {\it
sample 1} and {\it sample 2}, the best fits yield $\mu_1=-0.36$
($T_{90}\sim0.44$~s; $\sigma_1=0.46$) and $\mu_2=1.47$
($T_{90}\sim29.5$~s; $\sigma_2=0.47$) and  $\mu_1=-0.23$
($T_{90}\sim0.59$~s; $\sigma_1=0.22$) and $\mu_2=1.04$
($T_{90}\sim11$~s; $\sigma_2=0.57$), respectively. These
results show that the duration bimodality is not affected by the
spectral shape of GRBs, but the duration central value of
long bursts in {\it sample 2} is shifted to smaller value and the
duration distribution covers a smaller range (see the middle panel of
Figure 1). In other words, on average, the duration of long GRB with
power-law spectrum is smaller than that of GRB with curved
spectrum. It is well known that the duration strongly depends upon
the sensitivity and the energy range of the instrument. The bimodal
distribution of duration is either a robust result or a instrument
selection effect. Using the data of GRBs simultaneously detected by
Swift-BAT and Fermi-GBM, Virgili et al. (2011) found that the
duration distributions in the BAT and GBM bands are all bimodal. It
is essential to make a more detailed study explaining the fact that
the bimodal duration distribution is not presented in observations
of HETE-2 and Swift. The future work is going to explore it.
We also adopt the conventional $T_{90} = 2$ s to separate short and
long GRBs.

The spectral hardness is an additional discriminator for the
classification of GRBs. In this work, we firstly explore the
correlation between $HR$ and $T_{90}$, where $HR$ defines the
fluence ratio between two broad energy bands. We simply
divide the GBM energy band (10-1000 keV) into five energy bands,
i.e. 10-25, 25-50, 50-100, 100-300, 300-1000 keV. Using the
spectral data in N11, we are able to calculate $HR$ in arbitrary two
energy bands. We define four different $HR$ measurements, namely
{\it HR1}, between the 50-100 keV and the 25-50 keV energy bands,
i.e. the typical BAT energy bands; {\it HR2}, between the 100-300
keV and the 50-100 keV energy bands, i.e. the typical BATSE energy
bands; {\it HR3}, between the 300-1000 keV and the 10-300 keV energy
bands; and {\it HR4}, between the 100-1000 keV and the 25-100 keV
energy bands.

We find that the values of $HR$ are significantly different
in these energy bands. This is expected, since they strongly depend
on the energy bands between which they are calculated. Figure 2
shows $HR$ versus $T_{90}$ for 425 GRBs from N11. From this figure,
we find that there is an obvious tendency that short GRBs have
harder spectra than long GRBs. The correlations between $HR$ and
$T_{90}$ are reported in Table 1. We find that (1) for all
425 GRBs, $HR$ and $T_{90}$ are all anti-correlated for all choices
of $HR$ ({\it HR1-HR4}), but the correlation coefficients as well
as the slopes are different; (2) the values of $HR$ are larger and
the $HR - T_{90}$ anti-correlation is stronger in the typical BATSE
energy bands (the median value of $HR$ is 2.21, the correlation
coefficient is $r=-0.41$, the chance probability is
$p=2.8\times10^{-18}$, and the slope is $b=-0.12$) than those in the
BAT energy bands (the median value of $HR$ is 1.56, $r=-0.28$,
$p=5.2\times10^{-9}$, and $b=-0.06$); (3) if we consider
short and long GRBs separately, the correlations are very weak or
even negligible.

In addition, the question arises whether the $HR - T_{90}$
anti-correlation of GRBs depends on spectral shapes. For a
comparison with the previous results from the BASTE and BAT
observations, the correlations between $HR1$ and $T_{90}$,
and between $HR2$ and $T_{90}$ for {\it sample 1} and {\it sample 2}
are investigated. The results are shown in Figure 3 and Table 1.
We find that the anti-correlation between $HR$ and $T_{90}$
indeed depends on the spectral shape of GRBs, and the GRBs with the
curved spectra have a more clear correlation than those with the
power-law spectra in the same energy bands (for a detailed analysis,
see Table 1). Now, the fact that there is no obvious correlation
between $HR$ and $T_{90}$ in the Swift-BAT sample as well as other
samples detected by instruments with a narrow/softer energy bandpass
such as HETE-2 (Sakamoto et al. 2011) can be easily
understood. This is because (1) the $HR - T_{90}$ anti-correlation
depends upon the energy bands and the correlation in the typical BAT
energy bands is weak; (2) the majority of spectra of BAT
GRBs are best fitted by a single power-law model. This is due to the
fact that BAT only covers a narrow energy band. The $HR - T_{90}$
anti-correlation for GRBs with the power-law spectra is not
obvious; (3) for a given GRB sample, the values of $HR$ are small
and the values of $T_{90}$ are large for BAT-like softer detector,
this makes the $HR - T_{90}$ anti-correlation unclear.

The peak energy and spectral index are also usually adopted to
depict the hardness of GRBs. N11 found that, on average,
short GRBs have a higher peak energy ($\sim$ 490 keV) and a harder
low-energy spectral index ($\sim -0.50$) than long GRBs ($\sim$ 160
keV and $\sim -0.92$). We analyze the correlations between
$E_{\mathrm{peak}}$ (peak energy in the time integrated spectra) and
$T_{90}$, and between $E^{p}_{\mathrm{peak}}$ (peak energy in the
peak flux spectra) and $T_{90}$ for {\it sample 1}, and
between $\alpha$ (which represents the spectral index of GRBs best
fitted with a power-law model) and $T_{90}$ for {\it sample 2}. The
results are shown in Figures 4 and 5. We find that
$E_{\mathrm{peak}}$ and $T_{90}$, $E^{p}_{\mathrm{peak}}$ and
$T_{90}$, $\alpha$ and $T_{90}$ are all anti-correlated. The
detailed correlation analysis are presented in Table 2. The
fact that short GRBs have harder spectra and long GRBs have softer
spectra is confirmed once again. Recently, Gruber et al. (2011)
suggested that the rest fame peak energy distributions might be the
same for the two classes of GRBs based on a small Fermi GRB sample
with measured redshift. The different redshift distribution
of the two classes of GRBs might be responsible for the different
distributions of the observed peak energies (see e.g, Guetta \&
Piran 2005). However, we also should note that the redshift
distribution which has been found to be different for long and short
GRBs might have been strongly affected by the measurement methods.
Short GRBs tend to have lower redshift, very similar to those of
long GRBs measured by the same method, i.e., spectral analysis of
the presumed host galaxies (Shao et al. 2011). Therefore, the
distribution of the intrinsic peak energy of the two classes of GRBs
and the $E_{\mathrm{peak}}$ - $T_{90}$ anti-correlation are needed
to a further confirm. Moreover, for the short and long classes
separately, the $E_{\mathrm{peak}}-T_{90}$, $E^{p}_{\mathrm{peak}}-
T_{90}$, and $\alpha$ - $T_{90}$ anti-correlations do not exist or
even are positive correlated. These results are consistent
with what was found from the analysis of the correlation between
$HR$ and $T_{90}$. Our results indicate that hardness ratio,
peak energy and spectral index might be linked to the same physical
feature of GRBs. The nature is unknown so far.

\subsection{Distribution of Energy Ratio}

Although the distributions of hardness and duration can be used to
classify GRBs, the overlap between these two classes of GRBs can not
be ignored. Moreover, this scheme strongly relies on the subjective
choices required for the duration calculation. Recently,
Goldstein et al. (2010) showed that the $E_{\mathrm{peak}}$/Fluence
energy ratio (which physically represents a ratio of the energy at
which most of the gamma rays are emitted to the total energy emitted
in gamma rays) can be used as a new GRB classification
discriminator. This also has the big advantage that it does not rely
on the burst duration estimate. Using the preliminary duration and
spectral result of 382 Fermi-GBM GRBs, Goldstein et al. (2011)
analyzed the distribution of energy ratio and found that the
distribution separated into long bursts and short bursts
well. This supports the original claim obtained by Goldstein et al.
(2010). Meanwhile, we also analyze the distribution of the energy
ratio for {\it sample 1}. The fluence in the 10$-$1000 keV energy
band is calculated by using the spectral parameters in N11, then we
obtain the values of $E_{\mathrm{peak}}$/Fluence for each of GRBs in
{\it sample 1}. Figure 6 shows the distribution of the energy ratio.
In the top panel, we plot the 316 GRBs included in our {\it sample
1}. A bimodal distribution is evident, as previously shown by
Goldstein et al. (2011). By using a standard nonlinear
least-squares fitting algorithm, we fit the distribution by two
log-normal functions. The best fits yield a central value
$\mu_1=-1.34$ with a standard deviation $\sigma_1=0.62$, and
$\mu_2=0.27$ with $\sigma_2=0.35$, respectively. In the
bottom panel of Figure 6, we present two distributions corresponding
to long (unfilled histogram) and short (filled histogram) GRBs. This
further confirms the fact that (1) the bimodality of the energy
ratio distribution is correlated to that of the burst duration, and
that (2) the energy ratio is indeed a good discriminator for
classifying GRBs. Indeed, it could be used to identify some of the
controversial GBRs previously discussed (see Section 3.3). The
energy ratio bimodality can also be easily understood if we note
that short GRBs tend to have larger peak energies and smaller
fluences (due to their short durations) with respect to the long
ones.

To further check the differences between long and short
GRBs, we investigate the correlations between the peak flux ($P$)
and $T_{90}$ and between the $E^{p}_{\mathrm{peak}}/P$ ratio (which
physically represents a ratio of the energy at which most of the
gamma rays are emitted to the total energy emitted in gamma rays in
one second at peak of a burst) and $T_{90}$. Those two quantities
namely do not depend on the duration of a burst. The detailed
results are presented in Figure 7 and Table 2. Both the
correlations between $P$ and $T_{90}$ and between
$E^{p}_{\mathrm{peak}}/P$ and $T_{90}$ (shown in the top and bottom
panel of Figure 7, respectively) are different for long and short
GRBs. An anti-correlation between $P$ and $T_{90}$ and a positive
correlation between $E^{p}_{\mathrm{peak}}/P$ and $T_{90}$ are found
for all 234 GRBs with the peak flux curved spectra, although the
correlations are not very strong. Likewise, if we consider short and
long classes separately, the correlations are negligible or even
reversed. This further confirm that the correlation between the
hardness and the duration of GRBs is only a general trend between
two clusters of GRBs or two types of GRBs and does not apply to
either type.

\subsection{Intrinsic Spectral Energy Correlation}

Amati et al. (2002) found a tight correlation between the rest frame
peak energy, $E_{\mathrm{p,rest}}$ and the isotropic equivalent
gamma-ray energy, $E_{\mathrm{iso}}$, which was confirmed by later
observations (Amati 2006, 2010). However, short GRBs do not follow
the correlation, as is true for the peculiarly sub-energetic and
close GRB 980425, the proto-type of the GRB/SN connection (e.g.,
Amati 2006, 2010; Piranomonte et al. 2008; Ghirlanda et al. 2009,
Gruber et al. 2011). These facts suggest that the
$E_{\mathrm{p,rest}} - E_{\mathrm{iso}}$ plane may be used to
distinguish between different classes of GRBs and to understand the
differences in the physics/geometry of their emission. Here we
reanalyze this correlation taking into account the new observational
data. Only 7 short GRBs with redshift, reliable estimate of
$E_{\mathrm{peak}}$ and other spectral parameters are available by
the end of May 2011, as listed in Table 2.

The three peculiar/controversial short GRBs (GRBs 071227, 090927 and
100816A; Piranomonte et al. 2008; Amati et al. 2010; Gruber et al.
2011) are excluded for the following reasons. Caito et al. (2010)
suggested that GRB 071227 represents another example of a disguised
short GRB, after GRB 970228 and GRB 060614, on the basis of their
analysis performed in the context of the fireshell scenario. GRB
090927 and GRB 100816A are two short GRBs detected by Fermi and had
been analyzed by Gruber et al. (2011). However the spectrum of GRB
090927 is adequately fitted by a simple power law function (see,
Gruber et al. 2009; Nava et al. 2011), which has been confirmed by
our current analysis. The category of this GRB hence can not be
further identified due to its inaccurate peak energy measurement.
GRB 100816A was simultaneously detected by Swift-BAT, Fermi-GBM and
Konus-Wind, and its duration estimated by these three missions is
2.9$\pm0.6$ s (15-350 keV), 2 s (50-300 keV) and $\sim$2.8 s (20
keV-2 MeV), respectively. So it is difficult to determine
whether this GRB belongs to the short or to the long class only
based on its duration. Fan \& Wei (2011) identified a possible
wind-like medium surrounding this burst, which suggested a massive
star origin (i.e, this short-like GRB should be a long one). 
Moreover, this GRB does not deviate from the
$E_{\mathrm{p,rest}}-E_{\mathrm{iso}}$ region of long GRBs (see also
Gruber et al. 2011). We calculate the energy ratio and obtain
log($E_{\mathrm{peak}}$/Fluence)=-1.24 for GRB 100816A, which is a
typical value of long GRBs. Therefore, this event should belong to
the long class and has been included in our long GRB sample.

We also collected GRBs observed by Fermi-GBM with the same standard
as short GRBs analyzed above from the GCN. 27 GRBs
(including two short ones listed in Table 2) are obtained, as
listed in Table 3. Using the same calculation method as Amati et al.
(2008), we obtain their isotropic equivalent energies in the energy
range 1-10000 keV (in the bursts' rest frame). Figure 8
shows the correlation between $E_{\mathrm{p,rest}}$ and
$E_{\mathrm{iso}}$. Data reported in Amati et al. (2010) and the
references therein are also included. We find that the short GRBs
are significantly different from the long ones. To obtain a
quantitative comparison, we fit the
$E_{\mathrm{p,rest}}-E_{\mathrm{iso}}$ correlations for the short
and long GRBs separately. The best fits yield
\[E_{\mathrm{p,rest}}=2455\times
(\frac{E_{\mathrm{iso}}}{10^{52}})^{0.59\pm0.04}\ \mathrm{for \
short \ GRBs},\] with a spearman's rank correlation coefficient
$r=0.89$ and a chance probability $P=6.8\times$10$^{-3}$, and
\[E_{\mathrm{p,rest}}=100\times (\frac{E_{\mathrm{iso}}}{10^{52}})^{0.51\pm0.03}\ \mathrm{for \ long \ GRBs},\]
with $r=0.85$ and $P=1.2\times$10$^{-31}$ (see Table 1).  For long
GRBs, the result is in agreement with that obtained by some authors
(e.g., Amati 2010;  Ghirlanda et al. 2009; 2010; Gruber et al.
2011).

Wei $\&$ Gao (2003) discovered a tight correlation between the rest
frame peak energy and the luminosity ($L_{\mathrm{iso}}$) based on
nine GRBs (see their Fig.6). Such a correlation was soon confirmed by others
(e.g., Yonetoku et al. 2004; Ghirlanda et al. 2005).
Recently, further studies showed that although short GRBs
are inconsistent with the $E_{\mathrm{p,rest}}-E_{\mathrm{iso}}$
correlation hold by long GRBs, they might follow the
$E_{\mathrm{p,rest}}-L_{\mathrm{iso}}$ correlation (e.g., Ghirlanda
et al. 2009; Gruber et al. 2011). Such a trend has been
confirmed by our analysis with the latest data\footnote{The long GRB
data are taken form Ghirlanda et al. 2010 and the references
therein.} (see Figure 9). The fit to the
$E_{\mathrm{p,rest}}-L_{\mathrm{iso}}$ correlation yields
\[E_{\mathrm{p,rest}}=302\times (\frac{L_{\mathrm{iso}}}{10^{52}})^{0.40\pm0.03},\]
with $r=0.76$ and $P=2.3\times$10$^{-23}$. This might imply that the
energy dissipation processes powering the prompt gamma-ray emission
of short and long GRBs are rather similar though the progenitors
of these two kinds of events are likely different.

\section{SUMMARY}

In this work, we perform a statistical analysis of the temporal and
spectral properties of the latest Fermi-GBM GRBs catalogued by N11
to revisit the classification of GRBs. The traditional short/hard
and long/soft classification is supported for these Fermi GRBs by
the following facts: (1) The duration of the prompt gamma-ray
emission has a clear bimodal distribution. (2) The energy ratio
(i.e., $E_{\rm peak}/{\rm Fluence}$) also has a clear bimodal
distribution and its bimodality is correlated to that of the
duration distribution. (3) The anti-correlation between spectral
hardness (hardness ratio, peak energy, and spectral index) and
duration is also confirmed. Moreover, we find out that the
correlation between the hardness and the duration is only a general
trend between two clusters of GRBs or two types of GRBs and does not
apply to either type. Interestingly there are two leading models for
GRBs. One is the collapsar that likely produces long-lasting GRBs.
The other is the merger of two compact objects which is expected to
power short-living GRBs. However, it is not always reasonable to
argue a collapsar origin for long bursts and a merger origin for
short bursts since the possibilities that the collapsar can also
produce short GRBs and the merger can generate long GRBs can not be
ruled out, as already realized in the pre-Swift era (e.g.,
Klu{\'z}niak \& Ruderman 1998; Zhang et al. 2003; Fan et al. 2005).
The observations of some Swift GRBs, such as GRB 060614 and GRB
080503 (e.g., Xu et al. 2009), partly confirm such speculations.
Therefore additional discriminators are highly needed to classify
GRBs reliably.

In this work we also find that the anti-correlation between hardness
ratio and duration depends upon the spectral shapes of GRBs and the
energy bands. The bursts with the curved spectra in the typical
BATSE energy bands (between the 100-300 keV and the 50-100 keV
bands) show a tighter anti-correlation than those with the power-law
spectra in the typical BAT energy bands (between the 50-100 keV and
the 25-50 keV bands). This might explain why the $HR - T_{90}$
correlation is not evident for Swift-BAT GRB sample as well as other
GRB sample detected by instruments with a narrow/softer energy
bandpass such as HETE-2 (e.g., Sakamoto et al. 2011, Shao et al.
2011). We also analyze the intrinsic energy correlation for the
different GRB classes and find that all short GRBs deviate
significantly from the $E_{\mathrm{p,rest}}-E_{\mathrm{iso}}$
correlation hold by long GRBs, and they might follow another one
\[E_{\mathrm{p,rest}}=2455\times (E_{\mathrm{iso}}/10^{52})^{0.59}\]
based on the current small sample. The future observation will test
our result.

\acknowledgments

We are grateful to L. Nava for providing the published Fermi-GBM
data in their work. We acknowledge the GBM team for the public
distribution of the spectral properties of Fermi-GBM GRBs through
the GCN network. We also thank Dr. Yi-Zhong Fan for stimulating
discussion. This work was supported in part by the National Natural
Science Foundation of China (grants 10973041, 10921063, 11163003 and
11103083) and the National Basic Research Program of China
(No.~2007CB815404). F.-W.Z. acknowledges the support by the China
Postdoctoral Science Foundation funded project (No.~20110490139),
the Guangxi Natural Science Foundation (No.~2010GXNSFB013050) and
the doctoral research foundation of Guilin University of Technology.

\clearpage


\begin{figure}
\begin{center}
\includegraphics[width=1.0\textwidth]{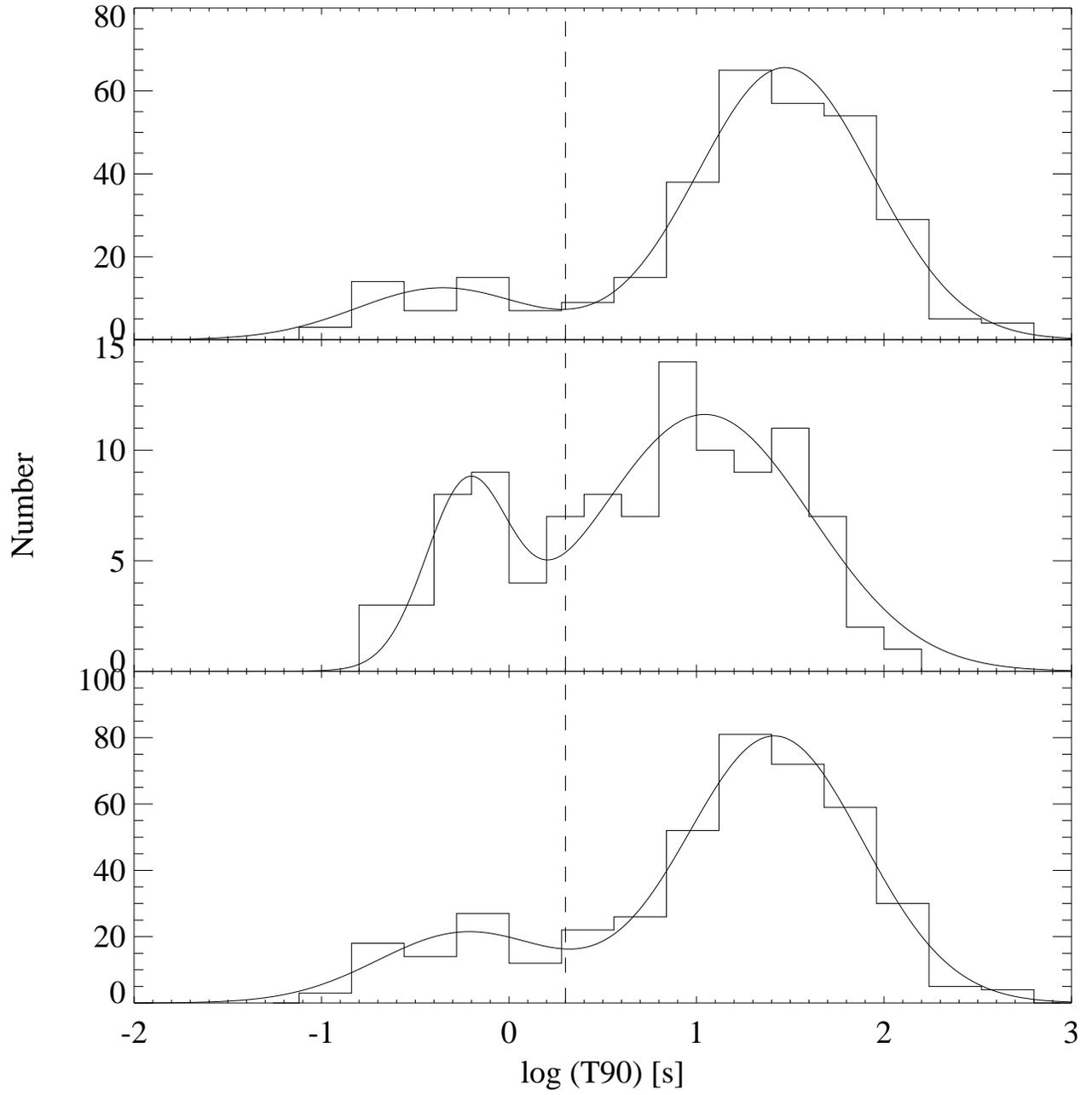}
\end{center}
\caption{Distributions of the durations for 322 GRBs  with the
curved spectra ({\it sample 1}, top panel), 103 GRBs  with the
power-law spectra ({\it sample 2}, middle panel), and all 425 GRBs
(bottom panel) observed by Fermi. The solid lines show the best fits
with two log-normal functions and the dashed vertical line is 2 s
separation line. The $T_{90}$ data are taken from Paciesas
et al. 2012.} \label{fig:lc}
\end{figure}

\begin{figure}
\begin{center}
\includegraphics[width=1.0\textwidth]{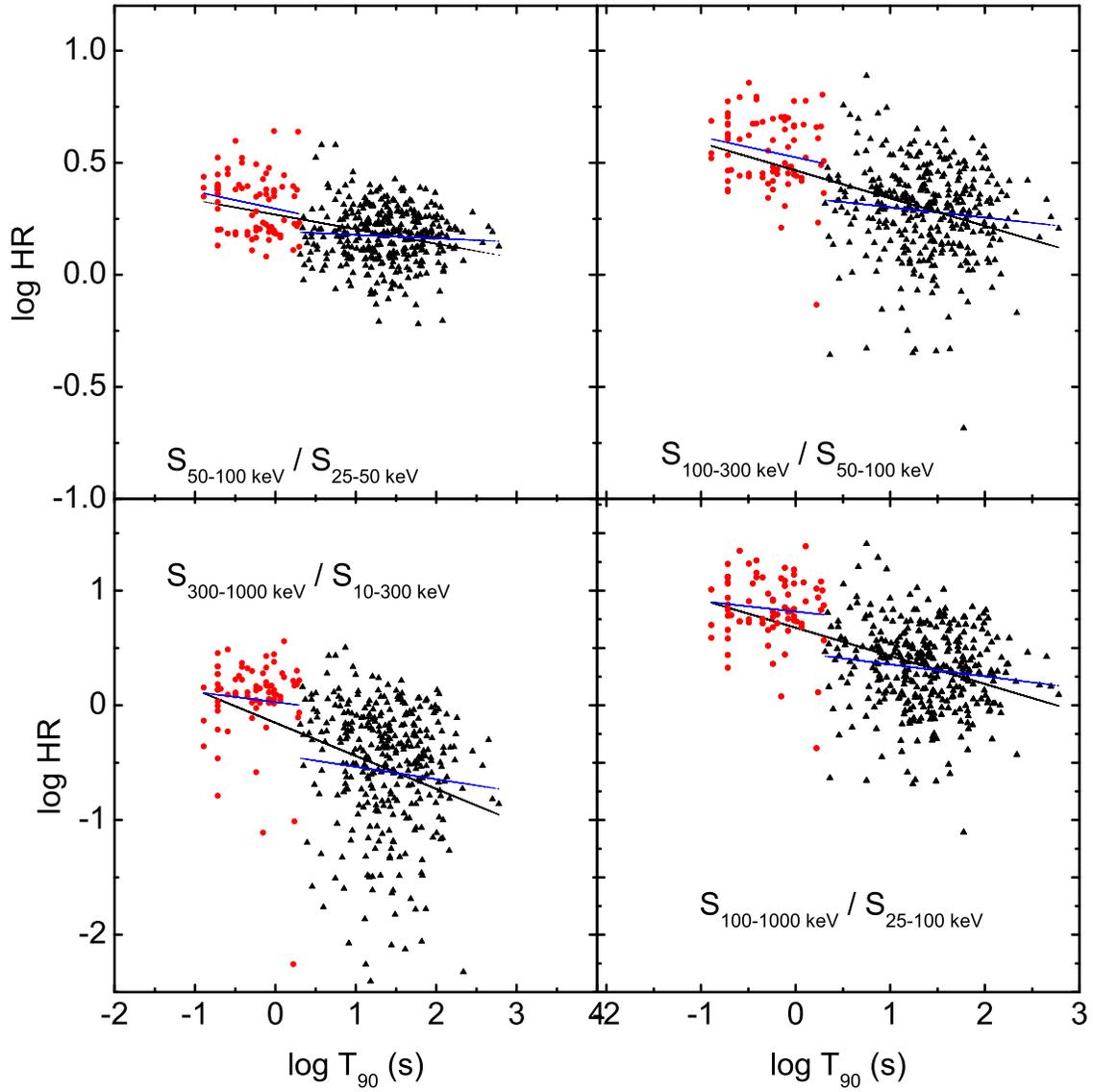}
\end{center}
\caption{Hardness ratio $\emph{versus}$ duration for 425 Fermi GRBs.
The filled circles represent the short GRBs and the filled triangles
represent the long ones. The solid lines are the best fits for all
bursts, the dashed lines are the best fits to short or long events
separately. For clarity, the estimated errors are not
shown.} \label{fig:hdrt90}
\end{figure}

\begin{figure}
\begin{center}
\includegraphics[width=1.0\textwidth]{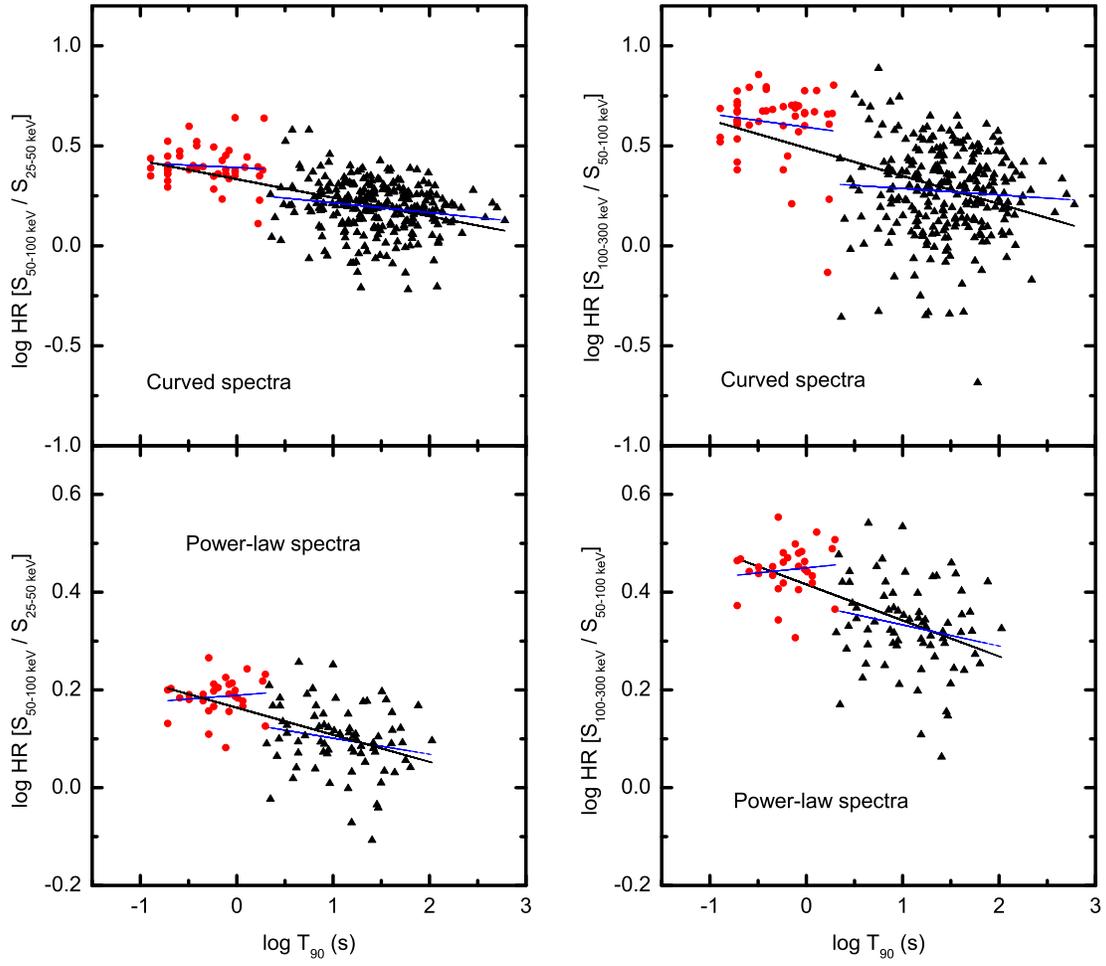}
\end{center}
\caption{The correlation between hardness ratio and duration for
GRBs with the different spectral shape. The other symbols are the
same as Figure 2.} \label{fig:hdrt90}
\end{figure}


\begin{figure}
\begin{center}
\includegraphics[width=0.8\textwidth]{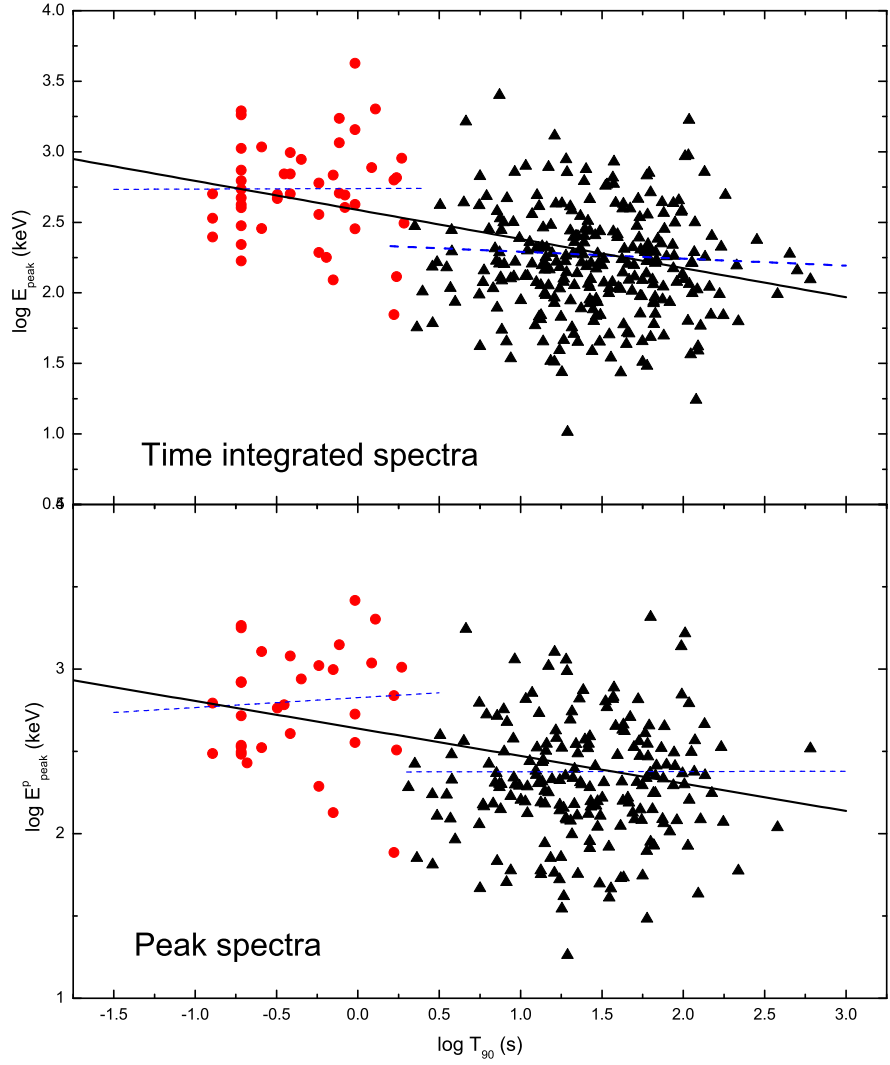}
\end{center}
\caption{$Top$: The correlation between the observed peak energies
in the time integrated spectra ($E_{\rm peak}$) and durations for
322 Fermi GRBs with the curved spectra. $Bottom$: The correlation
between the peak energies in the peak flux spectra ($E^{p}_{\rm
peak}$) and durations for 234 GRBs with the curved peak flux
spectra. The other symbols are the same as Figure 2.}
\label{fig:hdrt902}
\end{figure}

\begin{figure}
\begin{center}
\includegraphics[width=0.8\textwidth]{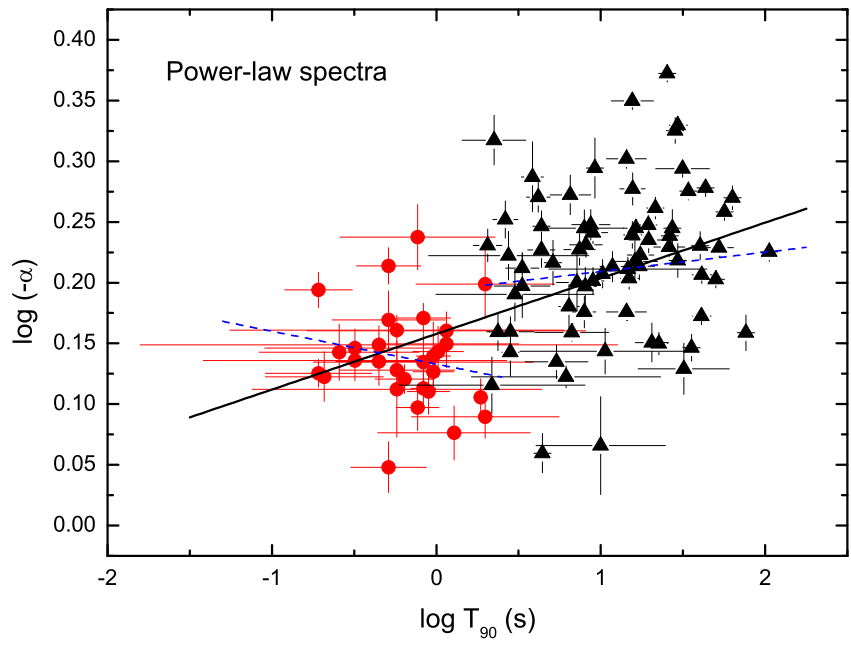}
\end{center}
\caption{Distribution of the spectral indices ($\alpha$)
$\emph{versus}$ durations for 103 GRBs with the power-law spectra.
The other symbols are the same as Figure 2.} \label{fig:hdr}
\end{figure}

\begin{figure}
\begin{center}
\includegraphics[width=0.8\textwidth]{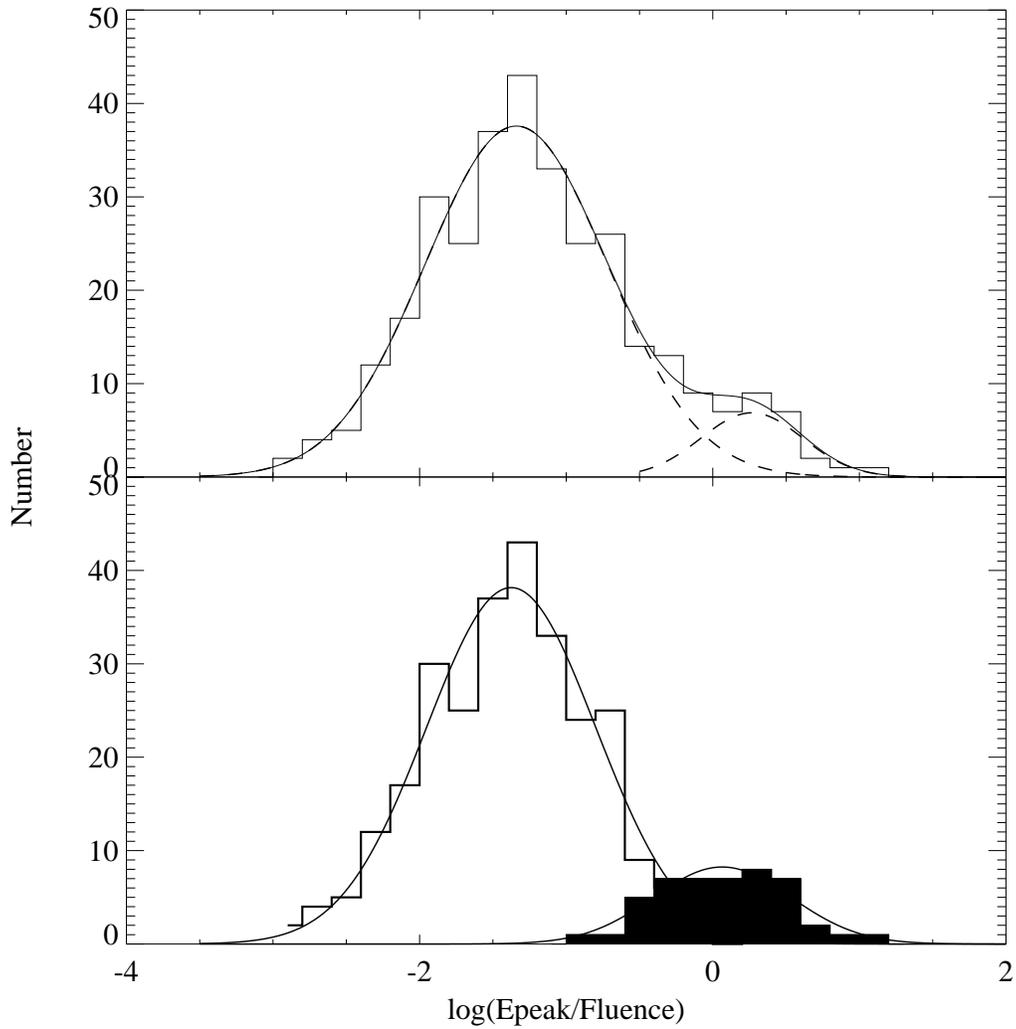}
\end{center}
\caption{Distribution of the energy ratio ($E_{\rm peak}$ /Fluence),
where the fluence is measured in the 10$-$1000 keV energy range. Top
panel shows the distribution for all 322 GRBs with the curved
spectra. The dashed lines are the best fits with two log-normal
functions separately and the solid line represent the superpositions
of two log-normal fitting. Bottom panel shows the distribution for
275 long GRBs (unfilled histogram) and 47 short GRBs (filled
histogram), where the solid curves are the best-fit log-normal
functions. There are clearly two distinct distributions. }
\label{fig:t90}
\end{figure}

\begin{figure}
\begin{center}
\includegraphics[width=0.8\textwidth]{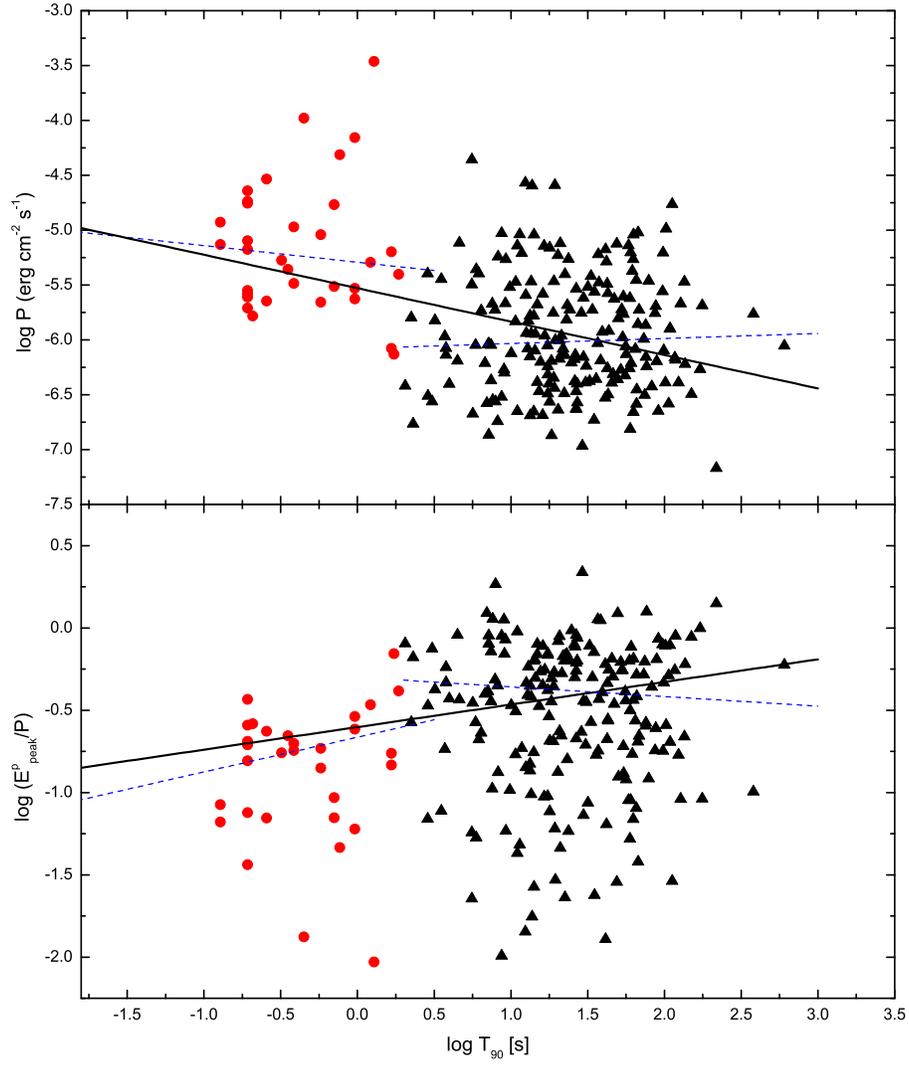}
\end{center}
\caption{The correlations between the peak flux ($P$) and the
duration ($T_{90}$), and between the $E^{p}_{\mathrm{peak}}/P$ ratio
and $T_{90}$ for 234 GRBs with the curved peak flux spectra, where
$E^{p}_{\mathrm{peak}}$ is the peak energy in the peak flux spectra
of a GRB. The other symbols are the same as Figure 2.}
\label{fig:redshift}
\end{figure}

\begin{figure}
\begin{center}
\includegraphics[width=1.0\textwidth]{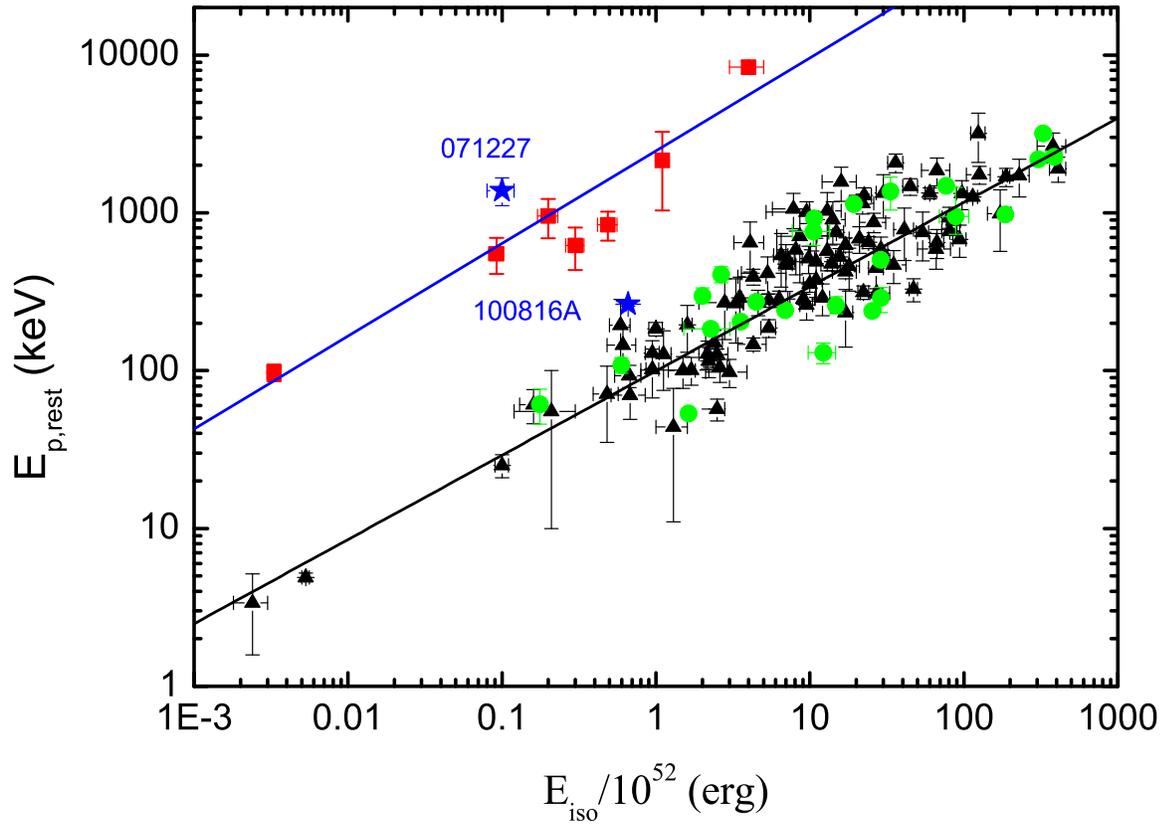}
\end{center}
\caption{The correlations between the rest frame peak energy
($E_{\mathrm{p,rest}}$) and the isotropic total energy
($E_{\mathrm{iso}}$). The squares represent the short GRBs,
the triangles represent Pre-Fermi long GRBs taken from Amati et al.
(2010) and the references therein, and the circles represent Fermi
long GRBs listed in Table 3. The stars are the two controversial
GRBs, GRB 071227 and GRB 100816A. The solid lines are the best fit
correlations: $E_{\mathrm{p,rest}}=2455\times
(\frac{E_{\mathrm{iso}}}{10^{52}})^{0.59}$ for short GRBs and
$E_{\mathrm{p,rest}}=100\times
(\frac{E_{\mathrm{iso}}}{10^{52}})^{0.51}$ for long GRBs.}
\label{fig:redshift}
\end{figure}


\begin{figure}
\begin{center}
\includegraphics[width=1.0\textwidth]{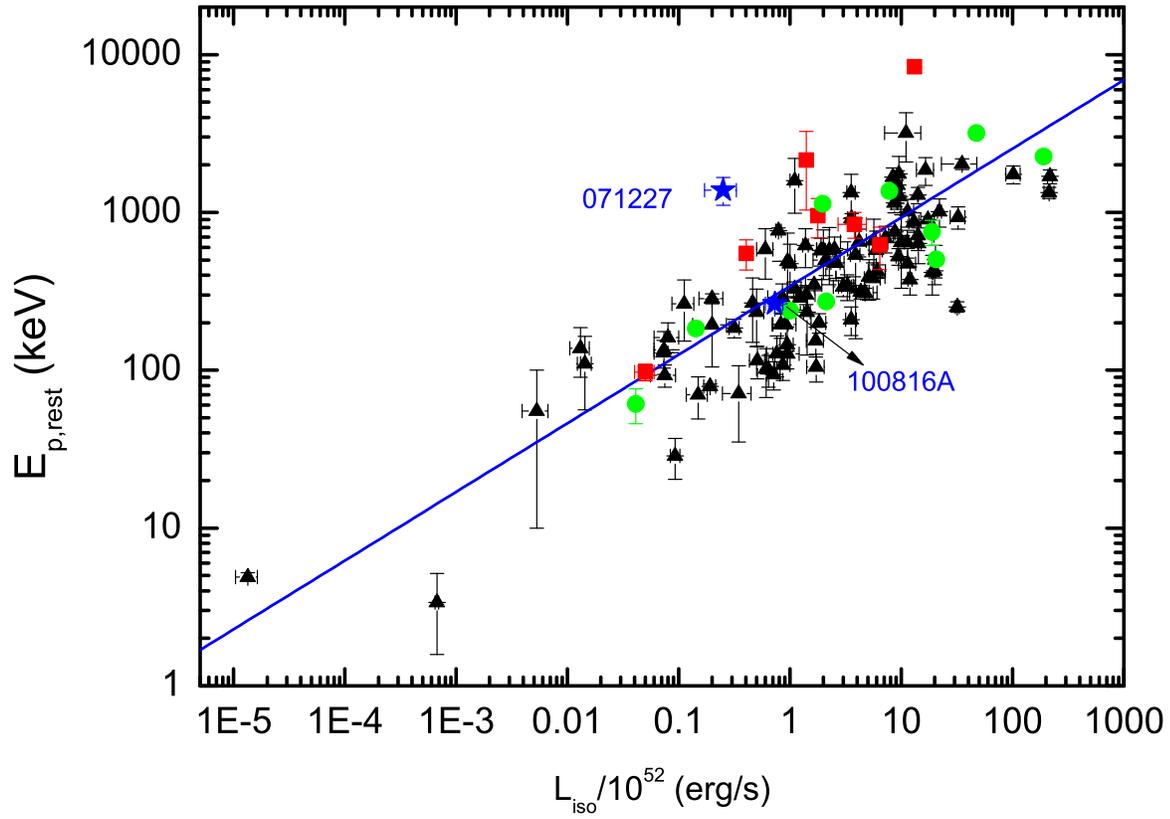}
\end{center}
\caption{The correlation between the rest frame peak energy
($E_{\mathrm{p,rest}}$) and the isotropic peak luminosity
($L_{\mathrm{iso}}$),  where the 105 long GRB data are taken from
Ghirlanda et al. 2010 and the references therein. The other symbols
are the same as Figure 8. Combined short and long GRBs, the best fit
to the $E_{\mathrm{p,rest}}-L_{\mathrm{iso}}$ correlation yield
$E_{\mathrm{p,rest}}=302\times
(\frac{L_{\mathrm{iso}}}{10^{52}})^{0.40\pm0.03}$.}
\label{fig:redshift}
\end{figure}

\clearpage

\begin{table*}
\small
\caption{Correlation analysis results.}
\label{short}
\begin{center}
\begin{tabular}{lllcclll}
\hline
\noalign{\smallskip}
Data$^{\star}$   & Number &  $HR^{\ast}$ &  $a$        &   $b$  &  $r$   &  Probability       \\
\noalign{\smallskip}
\hline
             &     &   &  &  log($HR$) = $a + b$ log($T_{90})$   \\
\hline
\noalign{\smallskip}
 All    & 425    & $HR1$             &      0.27$\pm$0.01   &  -0.06$\pm$0.01    & -0.28 & 5.2$\times$10$^{-9}$          \\
 All    & 425    & $HR2$            &      0.47$\pm$0.02   &  -0.12$\pm$0.01    & -0.41 & 2.8$\times$10$^{-18}$          \\
 All    & 425    & $HR3$               &      -0.15$\pm$0.05  &  -0.29$\pm$0.04    & -0.42 & 6.2$\times$10$^{-20}$          \\
 All    & 425    & $HR4$               &      0.68$\pm$0.03   &  -0.25$\pm$0.02    & -0.44 & 3.0$\times$10$^{-21}$          \\
 Short  & 77     & $HR1$           &      0.30$\pm$0.02  &  -0.08$\pm$0.04      & -0.22 & 0.06                          \\
 Short  & 77     & $HR2$         &      0.52$\pm$0.02  &  -0.09$\pm$0.05       & -0.13 & 0.25                         \\
 Short  & 77     & $HR3$            &      0.03$\pm$0.06  &  -0.09$\pm$0.13     & 0.04 & 0.74                          \\
 Short  & 77     & $HR4$            &      0.82$\pm$0.04  &  -0.09$\pm$0.10    & -0.03 & 0.81                          \\
 Long   & 348    & $HR1$            &      0.19$\pm$0.02  &  -0.02$\pm$0.01    & -0.03 & 0.56          \\
 Long   & 348    & $HR2$            &      0.35$\pm$0.03  &  -0.05$\pm$0.02    & -0.11 & 0.03           \\
 Long   & 348    & $HR3$             &      -0.43$\pm$0.11  &  -0.11$\pm$0.07   & -0.14 & 0.009                          \\
 Long   & 348    & $HR4$            &      0.46$\pm$0.06  &  -0.10$\pm$0.04    & -0.15 & 0.007           \\
\hline
 \noalign{\smallskip}
S1 All    & 322    & $HR1$            &      0.33$\pm$0.01  &  -0.09$\pm$0.01    & -0.44 & 8.4$\times$10$^{-17}$          \\
S2 All    & 103    & $HR1$          &      0.16$\pm$0.01  &  -0.05$\pm$0.01    & -0.55 & 1.5$\times$10$^{-9}$          \\
S1 All    & 322    & $HR2$          &      0.49$\pm$0.02  &  -0.14$\pm$0.02    & -0.36 & 2.2$\times$10$^{-11}$          \\
S2 All    & 103    & $HR2$          &      0.42$\pm$0.01  &  -0.07$\pm$0.01    & -0.55 & 1.5$\times$10$^{-9}$          \\
S1 Short  & 47     & $HR1$            &      0.39$\pm$0.02  &  -0.02$\pm$0.04    & -0.09 & 0.56                           \\
S2 Short  & 30     & $HR1$            &      0.19$\pm$0.01  &  0.02$\pm$0.03     & 0.16 & 0.40                           \\
S1 Short  & 47     & $HR2$           &      0.59$\pm$0.04  &  -0.07$\pm$0.07    & 0.05 & 0.72                           \\
S2 Short  & 30     & $HR2$           &      0.45$\pm$0.01  &  0.02$\pm$0.03     & 0.16 & 0.40                           \\
S1 Long   & 275    & $HR1$            &      0.26$\pm$0.03  &  -0.05$\pm$0.02    & -0.18 & 0.003           \\
S2 Long   & 73     & $HR1$            &      0.13$\pm$0.02  &  -0.03$\pm$0.02    & -0.20 &0.09           \\
S1 Long   & 275    & $HR2$           &      0.32$\pm$0.04  &  -0.03$\pm$0.03    & -0.07 & 0.23                           \\
S2 Long   & 73     & $HR2$           &      0.38$\pm$0.03  &  -0.04$\pm$0.02    & -0.20 &0.09           \\

\hline
\noalign{\smallskip}
\hline
\end{tabular}
\end{center}
$^{\star}$  S1 represents {\it sample 1} (316 GRBs with the curved spectra) and S2 represents {\it sample 2} (108 GRBs with the power-law spectra).\\
$^{\ast}$  The hardness ratios ($HR$) are measured between two
different energy bands, namely {\it HR1}, between the 50-100 keV and
the 25-50 keV energy bands, i.e. the typical BAT energy bands; {\it
HR2}, between the 100-300 keV and the 50-100 keV energy bands, i.e.
the typical BATSE energy bands; {\it HR3}, between the 300-1000 keV
and the 10-300 keV energy bands; and {\it HR4}, between the 100-1000
keV and the 25-100 keV energy bands.
\end{table*}

\clearpage

\begin{table*}
\small
\caption{Correlation analysis results.}
\label{short}
\begin{center}
\begin{tabular}{llccclll}
\hline
\noalign{\smallskip}
Data$^{\star}$   & Number &    $a$        &   $b$  &  $r$   &  Probability       \\
\noalign{\smallskip}
\hline

              &        &  &  log($E_{\mathrm{peak}}$) = $a + b$ log($T_{90}$)  \\
\hline
S1 All     & 322         &      2.59$\pm$0.03  &  -0.21$\pm$0.03    & -0.30   & 6.6$\times$10$^{-8}$  &         \\
S1 Short   & 47          &      2.74$\pm$0.07  &  0.01$\pm$0.13     & 0.08    & 0.59                     &         \\
S1 Long    & 275         &      2.34$\pm$0.08  &  -0.05$\pm$0.05    & -0.05   & 0.42                   &            \\
\hline
              &       &  & log($E_{\mathrm{peak}}^{\mathrm{p}}$) = $a + b$ log($T_{90}$)  \\
\hline
S3 All       & 234         &      2.64$\pm$0.04  &  -0.17$\pm$0.03    &-0.22 &7.0$\times$10$^{-4}$  &     \\
S3 Short     & 33          &      2.83$\pm$0.08  &  0.06$\pm$0.15     &0.14 & 0.42                     &          \\
S3 Long      & 201         &      2.37$\pm$0.09  &  0.01$\pm$0.06     &0.02 & 0.79                  &       \\
\hline
              &        &  & log($- \alpha$) = $a + b$ log($T_{90}$)  \\
\hline
S2 All       & 103          &      0.16$\pm$0.01  &  0.05$\pm$0.01     &0.55 &1.5$\times$10$^{-9}$ &     \\
S2 Short     & 30           &      0.13$\pm$0.01  &  -0.03$\pm$0.02    &-0.16 & 0.40               &          \\
S2 Long      & 73           &      0.19$\pm$0.02  &  0.02$\pm$0.02     &0.20 & 0.09                  &       \\
\hline
              &          &    & log($P$) = $a + b$ log($T_{90}$)  \\
\hline
S3 All       & 234          &      -5.53$\pm$0.05  &  -0.30$\pm$0.04    &-0.25  &9.0$\times$10$^{-5}$  &     \\
S3 Short     & 33           &      -5.29$\pm$0.12  &  -0.15$\pm$0.23    &-0.09 & 0.61                    &          \\
S3 Long      & 201          &      -6.08$\pm$0.11  &  0.05$\pm$0.07     &0.003 & 0.97                   &       \\
\hline
       &       &        & log($E_{\mathrm{peak}}^{\mathrm{p}}/P$) = $a + b$ log($T_{90}$)  \\
\hline
S3 All       & 234          &      -0.60$\pm$0.04  &  0.14$\pm$0.03     &0.18  & 0.006  &     \\
S3 Short     & 33           &      -0.66$\pm$0.08  &  0.21$\pm$0.16     &0.10 & 0.59                    &           \\
S3 Long      & 201          &      -0.30$\pm$0.09  &  -0.06$\pm$0.06    &0.01 & 0.85                   &       \\
\hline
       &       &        &  log($E_{\mathrm{p,rest}}$)= $a + b$ log($E_{\mathrm{iso}}$/10$^{52}$)   \\
\hline
Short   & 7     &      3.39$\pm$0.04  & 0.59$\pm$0.04    & 0.89 & 6.8$\times$10$^{-3}$   &           \\
Long    &110    &      2.00$\pm$0.01  & 0.51$\pm$0.03    & 0.85 & 1.2$\times$10$^{-31}$  &         \\
\hline
         &     &       & log($E_{\mathrm{p,rest}}$)= $a + b$ log($L_{\mathrm{iso}}$/10$^{52}$)   \\
\hline
         &112     &         2.48$\pm$0.03  & 0.40$\pm$0.03    & 0.76  & 2.3$\times$10$^{-23}$   &          \\

\noalign{\smallskip}
\hline
\end{tabular}
\end{center}
$^{\star}$ S1, S2 and S3 represent {\it sample 1} (322 GRBs with the
curved spectra), {\it sample 2} (103 GRBs with the power-law
spectra) and {\it sample 3} (234 GRBs with the curved peak flux
spectra), respectively.
\end{table*}

\renewcommand{\tabcolsep}{4pt}
\renewcommand{\baselinestretch}{0.5}
\begin{table*}

\tabletypesize{\small}
\caption{Short GRBs with measured redshifts and spectral parameters.}
\label{short}
\begin{center}
\begin{tabular}{lllllllllll}

\hline
\noalign{\smallskip}
GRB     & $z^{\star}$        &    $\alpha$   &  Peak flux & Range   &  $L_{\rm iso}$    &$E_{\mathrm{p,rest}}$&  Fluence     & range   &  $E_{\rm iso}$   &  Ref.  \\
        &          &               &            & (keV)   &  $10^{52}$ erg/s  &  (keV)       &  ($10^{-6}$) & (keV)     &  $10^{52}$ erg   &        \\
\noalign{\smallskip}
\hline
\noalign{\smallskip}
050709  &0.16      &-0.53$\pm$0.12    &  5.1$\pm$0.5E-6  & 2-400   &  0.05$\pm$0.01       & 97.4$\pm$11.6   &  0.4$\pm$0.04   & 2-400   &   0.0033$\pm$0.0001    &   1\\
051221  &0.5465    &-1.08$\pm$0.13    &  4.6$\pm$1.3E-5  & 20-2000 &  6.42$\pm$0.56       & 620$\pm$186     &  3.2$\pm$0.9    & 20-2000 &   0.3$\pm$0.04      &   1\\
061006  &0.4377    &-0.62$\pm$0.2     &  2.1e-5          & 20-2000 &  1.78$\pm$0.23       & 955$\pm$267     &  3.57        & 20-2000 &   0.2$\pm$0.03      &   1\\
070714  &0.92      &-0.86$\pm$0.1     &  2.8$\pm$0.3     & 100-1000&  1.4$\pm$0.1         & 2150$\pm$1113   &  3.7         & 15-2000 &   1.1$\pm$0.1       &   1\\
090510  &0.903    &-0.80$\pm$0.03      &   80            & 8-40000 &  13.1$\pm$0.87       & 8370$\pm$760    &  30$\pm$2       & 8-40000 &   3.75$\pm$0.25     &
2 \\
100117A  &0.92     &-0.14$^{+0.33}_{-0.27}$ &6.1$\pm$0.4   & 8-1000  &  0.41$\pm$0.05       & 551$^{+142}_{-96}$  &  0.41$\pm$0.05  & 8-1000 &  0.09$\pm$0.01 & 3\\
101219A &0.718    &-0.22$^{+0.30}_{-0.25}$ &2.8$\pm$0.8E-6 & 20-10000 & 3.78$\pm$1.08      & 842$^{+177}_{-136}$ & 3.6$\pm$0.5    & 20-10000 &  0.49$\pm$0.07 & 4\\
\hline
071227$^{\ast}$ &0.383     &-0.7                &3.5$\pm$1.1E-6     & 20-1300 &  0.25$\pm$0.08      & 1384$\pm$277      & 1.6$\pm$0.2    & 20-1300 &  0.1$\pm$0.02 & 5\\

\noalign{\smallskip}
\hline
\end{tabular}
\end{center}
 References: (1) Ghirlanda et al. (2008); (2) Guiriec et al. 2009 ; (3) Paciesas et al. 2010; (4) Golenetskii et al. 2010; (5) Golenetskii et al. 2007. \\
\textbf{Notes.} $^{\ast}$ Some authors suggested that this GRB is a disguised short one as GRB 060614 (e.g., Caito et al. 2010). \\
                $^{\star}$ The redshift are taken from the Greiner's webpage (http://www.mpe.mpg.de/~jcg/grbgen.html).

\end{table*}

\clearpage
\renewcommand{\arraystretch}{0.8}

{\small{
\begin{table*}

\caption{The spectral parameters of 27 Fermi-GBM GRBs with known redshifts and well measured peak energies. The redshift are taken from the Greiner's webpage.}
\tabletypesize{\small}
\tablewidth{0pt}
\label{short}
\begin{center}
\begin{tabular}{llcclllll}
\hline
\noalign{\smallskip}
GRB     & z        &    $\alpha$   & $\beta$  & $E_{\mathrm{peak}}$  & Fluence            &   range      &  $E_{\mathrm{iso}}$    & GCN   \\
        &          &               &          & (keV)       &  $(10^{-6}$ erg/$cm^{2}$)  &  (keV)       & ($10^{52}$ erg)   & number \\
\noalign{\smallskip}
\hline
   080810 &       3.35 &          -0.91$\pm$0.12 &         ... &           313.5$\pm$73.6 &        6.9$\pm$0.5 &    50-300  &   33.3$\pm$2.4 &       8100 \\

   080916A &      0.689 &       -0.9$\pm$0.1 &              ... &             109$\pm$9 &         15$\pm$5 &   25-1000  &    2.27$\pm$0.76 &       8263 \\

   080916C &       4.35 &      -0.91$\pm$0.02 &      -2.08$\pm$0.06 &        424$\pm$24 &        190 & 8-30000&   387$\pm$46 &       8278 \\

    081007 &     0.5295 &       -1.4$\pm$0.4 &             ... &         40$\pm$10 &        1.2$\pm$0.1 &    25-900  &    0.18$\pm$0.02 &       8369 \\

    081222 &       2.77 &      -0.55$\pm$0.07 &       -2.1$\pm$0.06 &        134$\pm$9 &       13.5$\pm$0.8 &    8-1000  &   28.8$\pm$1.7 &       8715 \\

    090323 &       3.57 &      -0.89$\pm$0.03 &            ... &        697$\pm$51 &        100$\pm$1 &    8-1000  &   327$\pm$3 &       9035 \\

    090328 &      0.736 &      -0.93$\pm$0.02 &       -2.2$\pm$0.1 &        653$\pm$45 &       80.9$\pm$1 &    8-1000  &   19.3$\pm$0.2&       9057 \\

    090423 &       8.26 &      -0.77$\pm$0.35 &           ... &         82$\pm$15 &        1.1$\pm$0.3 &    8-1000  &   10.6$\pm$2.9 &       9229 \\

    090424 &      0.544 &       -0.9$\pm$0.02 &       -2.9$\pm$0.1 &        177$\pm$3 &         52$\pm$1 &    8-1000  &    4.47$\pm$0.09 &       9230 \\

    090510$^{a}$ &      0.903 &       -0.8$\pm$0.03 &       -2.6$\pm$0.3 &       4400$\pm$400 &         30$\pm$2 & 8-40000 &    3.75$\pm$0.25 &       9336 \\

    090516 &      4.109 &      -1.51$\pm$0.11 &           ... &      185.6$^{+98.4}_{-42.5}$ &         23$\pm$5 &    8-1000  &   88.5$\pm$19.2 &       9415 \\

    090618 &       0.54 &      -1.26$^{+0.06}_{-0.02}$ &       -2.5$^{+0.15}_{-0.33}$ &      155.5$^{+11.1}_{-10.5}$  &        270$\pm$6 &     8-1000 &   25.4$\pm$0.6 &       9535 \\

    090902B &      1.822 &     -0.70$\pm$0.01 &      -3.85$^{+0.21}_{-0.31}$ &        775$\pm$11 &        374$\pm$3 & 50-10000 &   305$\pm$2 &       9866 \\

   090926A &     2.1062 &      -0.75$\pm$0.01 &      -2.59$^{+0.04}_{-0.05}$ &        314$\pm$4 &        145$\pm$4 &     8-1000 &   186$\pm$5 &       9933 \\

   090926B &       1.24 &      -0.13$\pm$0.06 &          ... &         91$\pm$2 &        8.7$\pm$0.3 & 10-1000    &    3.55$\pm$0.12&       9957 \\

   091003A &     0.8969 &      -1.13$\pm$0.01 &      -2.64$\pm$0.24 &      486.2$\pm$23.6 &       37.6$\pm$0.4 &     8-1000 &   10.6$\pm$0.1 &       9983 \\

    091020 &       1.71 &        -0.2$\pm$0.4 &       -1.7$\pm$0.02 &       47.9$\pm$7.1 &         10$\pm$2 &     8-1000 &   12.2$\pm$2.4 &      10095 \\

    091127 &       0.49 &      -1.27$\pm$0.06 &       -2.2$\pm$0.02 &         36$\pm$2 &       18.7$\pm$0.2 &     8-1000 &    1.63$\pm$0.02 &      10204 \\

   091208B &      1.063 &      -1.48$^{+0.05}_{-0.05}$ &          ... &      144.2$^{+18}_{-13.9}$  &        5.8$\pm$0.2 &     8-1000 &    2.01$\pm$0.07 &      10266 \\

   100117A$^{a}$ &       0.92 &      -0.14$^{+0.33 }_{-0.27}$ &           ... &        287$^{+74}_{-50}$  &       0.41$\pm$0.05 &     8-1000 &    0.09$\pm$0.01&      10345 \\

   100414A &      1.368 &      -0.58$\pm$0.01 &           ... &      627.6$^{+12.5}_{-12.1}$ &        129$\pm$2 &     8-1000 &   76.6$\pm$1.2 & 10595 \\

   100728B &      2.106 &       -0.9$\pm$0.1 &            ... &        131$\pm$15 &        2.4$\pm$0.1 &     8-1000 &    2.66$\pm$0.11 &  11015\\

   100814A &       1.44 &      -0.64$^{+0.14}_{-0.12}$ &      -2.02$^{+0.09}_{-0.12}$ &      106.4$^{+13.9}_{-12.6}$ &       19.8$\pm$0.6 &    10-1000 &   14.8$\pm$0.5 &      11099 \\

   100816A$^{b}$ &     0.8049 &      -0.31$\pm$0.05 &      -2.77$\pm$0.17 &      136.7$\pm$4.7 &       3.84$\pm$0.13 &    10-1000 &    0.73$\pm$0.02 &      11124 \\

   100906A &      1.727 &      -1.34$^{+0.08}_{-0.06}$ &      -1.98$^{+0.06}_{-0.07}$ &        106$^{+17.5}_{-20.2}$ &       26.4$\pm$0.3 &    10-1000 &   28.9$\pm$0.3 &      11248 \\

   101219B &       0.55 &       0.33$\pm$0.36 &      -2.12$\pm$0.12 &         70$\pm$8 &        5.5$\pm$0.4 &    10-1000 &    0.59$\pm$0.04 &      11477 \\

   110213A &       1.46 &      -1.44$\pm$0.05 &          ... &       98.4$^{+8.5}_{-6.9}$
 &       10.3$\pm$0.3 &    10-1000 &    6.9$\pm$0.2 &      11727 \\

\noalign{\smallskip}
\hline
\end{tabular}
\end{center}
\textbf{Notes.} $^{a}$ Two short GRBs. \\
$^{b}$ This GRB was classified as a short one by the previous study (e.g., Gruber et al. 2011), we find it should belong to the long class.

\end{table*}
}
}

\end{document}